\documentstyle[preprint,eqsecnum,aps,floats]{revtex}
\begin{document}
%
%
\input epsf
%

\draft
\preprint{UCSBTH-95-9,  NSF-ITP-96-06}
\date{\today}

\title{Scientific Knowledge from the \\
Perspective of Quantum Cosmology\footnote{Contribution to the
proceedings of the workshop {\sl Limits to Scientific Knowledge} held at
Abisko, Sweden, May 15-19, 1995}}

\author{James B.
Hartle\thanks {e-mail: hartle@cosmic.physics.ucsb.edu}}
\vskip .13 in
\address{Department of Physics and Institute for Theoretical Physics, \\
University of California,
Santa Barbara, CA 
93106-9530 USA}

\maketitle

\begin{abstract}
\tighten

Existing physical theories do not predict every feature of our
experience but only certain regularities of that experience. That
difference between what could be observed and what can be predicted is one
kind of limit on scientific knowledge.  Such limits are inevitable if
the world is complex and the laws governing the regularities of that
world are simple.  Another kind of limit on scientific knowledge
arises because even simple theories may
require intractable or impossible computations to yield specific
predictions. 
A third kind of limit concerns our ability to know theories through the
process of induction and test.  Quantum cosmology --- that part of
science concerned with the quantum origin of the universe and its
subsequent evolution --- displays all three kinds of limits. This paper
briefly describes quantum cosmology and discusses these limits. The place
of the other sciences in this most comprehensive of physical frameworks is
described.

\end{abstract}


\tighten
\narrowtext
\setcounter{footnote}{0}
\section{Introduction}
\label{sec:I}

The assignment of the organizers was to speak on the subject of ``limits
to scientific knowledge''. This is not a topic on which I have been
forced to reflect a great deal in the course of my efforts in
astrophysics, but I shall try to offer a few thoughts on it from the
perspective of cosmology. Like any assignment, the first task is to
understand what it might mean. I shall say more about this later, but
one thing is immediately clear: This is not simply an empirical
question, but rather concerns the relationship between what we observe and
our theories of what we observe. Limits therefore depend on theories and
will vary from one scientific theory to another. The question of what
are the {\it fundamental} limits to scientific knowledge must be examined in
the most general theoretical context. In physics this is the subject of
quantum cosmology --- the quantum mechanics of the universe as a whole
and everything inside it. The nature of scientific knowledge in this
most comprehensive of theories is the subject of this essay. I shall try to
describe a little of what quantum cosmology is about, and address the
question of limitations to scientific knowledge in this most general of
contexts. 

\section{Three Kinds of Limits to Scientific Knowledge}
\label{sec:II}

In this Section three different kinds of limits to  scientific knowledge
are identified.  No claim is made that these are the only kinds of
limits, but these three have a general character that is inherent in the
nature of the scientific enterprise. Subsequent sections will illustrate
these general kinds of limits with examples from quantum cosmology.

\subsection{Limits to What is Predicted}
\label{subsec:A}

The task of science, as Bohr said, is ``to extend the range of our
experience and to reduce it to order'' \cite{Boh34}. 
To reduce experience to order is to compress the length
of a description of that experience. That compression is achieved when
a computer program can be exhibited which, given certain input,
outputs a string describing some parts of our experience and the length
of that program together with its
input are shorter than the length of the
output description. Theory
supplies the program. For instance, a detailed description of the
observations of the positions of the planets over the last 100 years
might make up a very long table, but Newton's equations of motion can
be used to compress all that information into two much shorter strings:
a string  
stating Newton's theory and another string giving the positions
and velocities of the planets at one time.

It is a logical possibility that {\it every} feature of
our experience --- the wave function of every quark, the velocity of
every molecule,  the position of
every leaf, the character of each biological species, the action of
every human, etc. --- is just a very long output of a short computer
program with {\it no} input. However, in the history of
scientific inquiry there is no evidence that the universe is so regular. 
Even the most
deterministic classical theories did not claim this. 
With Newtonian mechanics, Laplace
proposed only to predict the future and retrodict the past
 {\it given} the present position and velocity of
each particle in the universe. That list of initial data would be vastly
longer than a few treatises on Newtonian mechanics.
Existing theories predict a string describing our experience only given
some other, shorter, string as input. Theories do not predict everything
that is observed, but only certain regularities in what is observed. 
Some things are
predicted, some are not, and that limit to what is predicted is one
kind of limit to scientific knowledge.

Scientific laws must have some degree of simplicity to be discoverable,
comprehensible, and effectively applicable by human beings and other
complex adaptive systems. If the
complexity of the present universe is large, then this necessary
simplicity of the
laws implies that this kind of limit to scientific knowledge is inevitable. 
Not everything
can be predicted but only those regularities that are 
summarized in the laws of
science. In the following we shall describe what is predicted and what
is not predicted in quantum cosmology.\footnote{For a lucid discussion
in popular language of the notion of complexity and
 of prediction in quantum cosmology, as well as a
summary of some of the author's work with M.~Gell-Mann, see 
\cite{Gel94}.} 

\subsection{Limits to Implementation}
\label{sec:B} 

To be tested, the predictions of an abstractly represented theory
covering a broad class of phenomena must be implemented in particular
circumstances. The theory must produce numbers, and that process involves
computation. 
Even
if the laws are precisely specified, even if the input to those laws is
exactly stated, limitations of our ability to compute may limit our
ability to predict. This is another kind of limit to scientific
knowledge. The practical limitations of present computing machines
are all too familiar. Computing the motion of every particle in a
classical gas  $10^{22}$ particles in less than its real evolution
time is well
beyond the powers of contemporary computers. However, beyond the
limitations of contemporary machinery, we may ask whether
there are fundamental limitations on what can be computed that are
inherent in the form of the laws
themselves. The phenomenon of chaos is the source of one kind of
limitation. The precision
required of initial data to extrapolate a given time into the future
increases exponentially with that time for a wide variety of classical
systems. Another kind of limit arises in
cosmology where resources for computation, both in time and space,
are limited. Further, as we shall see, there is some evidence that
certain predictions of quantum cosmology may be non-computable numbers.

It is not difficult to display predictions which are computationally
intractable but which are measurably inaccessible. Given initial
conditions, classical theory predicts the orbits of every molecule of gas
in a room.  The explicit computation of this prediction at the operating
speeds of present computers would take much longer than the age of the
universe because of the large number of particles involved. Yet,
for the same reason, neither the initial condition nor the 
predicted orbits are measurably accessible quantities.
Merely exhibiting phenomena which are impossible or intractable to
compute is not much of a limit if the phenomena are impossible or
extraordinarily difficult to measure. The most interesting limits
concern
phenomena that are easy to measure but difficult to compute.

\subsection{Limits to Verification}
\label{sec:C}

The above discussion has assumed that we know the laws of physics. 
However, we arrive at those laws by a process of induction and test. 
Competing laws consistent
with known regularities are winnowed by the process of
checking their predictions with new observations. Are there fundamental
limits to what we can test, and therefore fundamental limits to how well
the theory can be known? Cosmology will provide examples.

\subsection{False Limits}
\label{sec:D}

Beware of false limits that arise only from imprecise language or the
comparison of a correct theory with an incorrect one. A
classic example is provided by the uncertainty principle in quantum mechanics
\begin{equation}
\Delta x\, \Delta p \geq \hbar/2\ .
\label{twoone}
\end{equation}
That relation is sometimes described as a limit on our ability to
predict (or ``measure'' or ``know'') both the position and momentum of a
particle at one time
 to accuracies better than those restricted by (\ref{twoone}).
However, the uncertainty principle 
 is more accurately characterized as a limit
on the use of classical language in a
quantum mechanical situation.

There is no state of a quantum mechanical particle with a precisely
defined position and momentum. That is the content of (\ref{twoone}).
The uncertainty principle, therefore,
is not a limit to what observed properties of
quantum particle are predicted by the theory. Since there is no quantum
state with precisely defined position and momentum, quantum theory
predicts that we shall never observe both simultaneously. 
Thus, as far as the
position and momentum of a particle are concerned, there is no disparity
within quantum theory
between what can be predicted and what is observed arising from
(\ref{twoone}) as there would be in
the case of a genuine limit of the type discussed in Section A.

As mentioned earlier, limits to prediction are properties of the
theories which specify what can be predicted. Of course, if we {\it
compare} two theories one may predict different phenomena from the
other. In classical physics there are states in which
the position and momentum of a particle {\it
are} simultaneously specified. In quantum theory there are not.
But quantum theory is correct
and classical theory incorrect for the domain of phenomena we have in
mind. The uncertainty principle (\ref{twoone}) may be viewed as a kind of
limit to how far classical concepts and language can be applied in
quantum theory, but, were we to strictly adhere to the language and
concepts of quantum theory, it would be no limit at all.

\section{Dynamical Laws and Initial Conditions}
\label{sec:III}

As we mentioned above, fundamental limits 
to what is predicted, to how predictions can
be implemented, and to how theory can be verified depend on what
the basic theory is. This Section sketches some essential
features of basic physical theory today. Of course, we are on
dangerous ground here.  The most basic laws are often the
furthest from definitive experimental test. Nevertheless, it is
interesting to see what kinds of limits might exist in the kind of basic 
theoretical framework that is under active investigation by physicists today. 

The most general framework for prediction is quantum cosmology --- the
quantum theory of the universe as a whole and everything that goes on
inside it. In the following I shall
describe a little of this theory. 

Historically, physics for the most part has been concerned with finding
dynamical laws --- laws which compress the description of evolution over
time to the description of an initial condition. Thus, these
dynamical laws require
boundary conditions to yield predictions. There are no particular laws
governing these boundary conditions. They are specified by our
observations of the part of the universe outside the subsystem whose
dynamics is of interest. In a room,
 if we observe no incoming radiation, we
solve Maxwell's equations there with no-incoming-radiation boundary
conditions. If we prepare an atom in a certain atomic state we solve
Schr\"odinger's equation with that initial condition, etc.

But in cosmology we are confronted with a fundamentally
different kind of problem. Whether classical or quantum, the
dynamical laws governing the evolution of the universe 
require boundary conditions. But in
cosmology there is no
``rest of the universe'' to pass their specification off to. The boundary
conditions must be part of the laws of physics themselves. There is no
other place to turn.

A present view, therefore, 
 is that the most general laws of physics involve two elements:

\begin{itemize}
\item  The laws of dynamics prescribing the evolution of matter and
fields and consisting of a unified theory of
the strong, electromagnetic, weak, and gravitational forces.
\item  A law specifying the initial boundary condition of the universe.
\end{itemize}
There are no predictions of any kind which do not depend on these two
laws, even if only very weakly, or even when expressed through
phenomenological approximations to these laws (like classical physics) 
appropriate in particular and limited
circumstances with forms that may be only distantly
related to those of the basic theory.

The search for a fundamental theory of the dynamics of matter has been
seriously under way since the time of Newton. Classical mechanics,
Newtonian gravity, electrodynamics, special relativity, general
relativity, quantum mechanics, quantum
electrodynamics, the theory of the electroweak
interactions, quantum chromodynamics, grand unified theories, and 
superstring theory are but some of the important milestones in this
search. The search for a theory of the initial condition of the universe
has been seriously under way for not much more than a decade. (See
Ref.~\cite{Hal91} for a review.) The reason
for this difference can be traced to the scales on which the regularities
summarized by these two laws emerge. The trajectory of a ball in the air,
the flow of water in a pipe, or the motion of a planet in the solar system
all exhibit the regularities implied by Newtonian mechanics. The
regularities of the dynamical laws of atomic and particle physics can be
exhibited in
experiments carried out in laboratories or large accelerators. However, 
characteristic regularities implied by a theory of the initial condition 
of the universe emerge mostly
on much larger, cosmological scales.
\begin{figure}[t]
\centerline{\epsfysize=2.50in \epsfbox{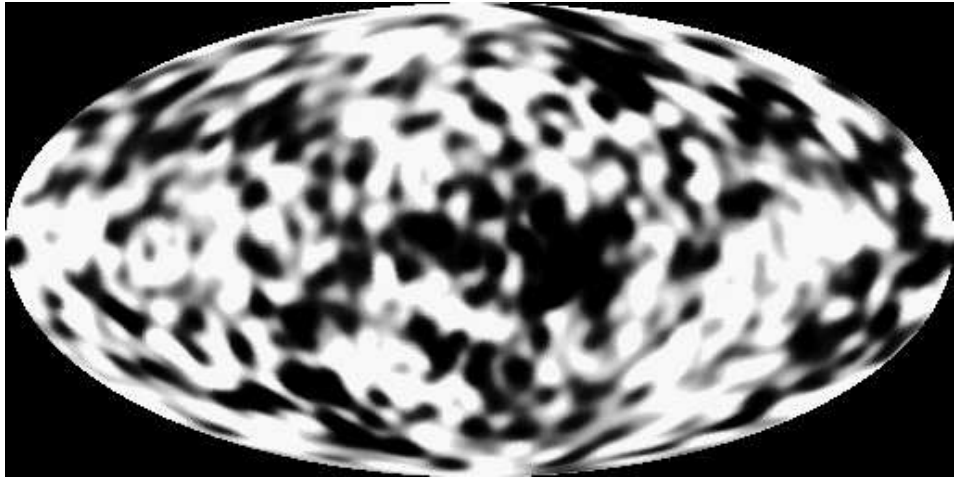}}
\vskip .13 in
\caption
{\sl A sky map of temperature fluctuations in the cosmic background
radiation.
This figure may be thought of as a picture of the universe approximately
300,000 years after the big bang. The hot mixture of matter and
radiation that exists immediately after the big bang cools as the
universe expands.
About 300,000 years later the universe has cooled enough that matter
and radiation no longer significantly interact. Photons from that time
have been traveling freely towards us ever since. Their characteristic
temperature now is only 2.7 degrees above absolute zero, yet they can be
detected at microwave wavelengths by sensitive
instruments. The figure
above shows  a sky map of the temperature of that radiation based on
data taken with the COBE satellite. The dark spots are where the
sky is cooler than the mean temperature and the white areas are where it
is hotter.  The differences in temperature
between the darkest black and the whitest while it is only a few hundred
{\it micro} degrees Kelvin. The universe is thus essentially featureless
at 300,000 years after the big bang except for these tiny fluctuations.
These small fluctuations, however, are the origin of all the complexity
in the universe that we see today. [Greyscale adaptation by J.~Gundersen
of the results of C.~Bennett, et.~al.~{\sl Ap.~J.}, {\bf 436}, 423
(1994)]}
\end{figure}

On any scale the universe exhibits some regularities in space as
distinct from regularities in time. Rocks on
one part of the earth are related to rocks on another part. 
Similarly, there are relations between individual members of 
 biological species, and human history in different locations. 
These regularities
have their origins in the common origin of rocks in the earth,
the evolution of biological species, and the facts of human history. 
On cosmological scales the universe is more regular in space than it is on
smaller scales.      
The progress of observation in astronomy in recent decades has given us an
increasingly detailed picture of the universe on ever larger scales of
space and time. The remarkable inference from these observations
is that the universe becomes
increasingly simple as we move to larger scales
in space and more distant times in the past.
 Galaxies are not very complicated objects but
still exhibit a variety of types and considerable
individuality. On the larger scale of a tenth of the radius of the universe
the galaxies are no longer individual objects, but there is 
considerable structure in their distribution. Pictures
of the distribution of galaxies
on the sky, which probe out to greater distances show less structure.
On the largest scales,
the distribution of the cosmic background radiation temperature,
which is as close as we can come to a picture of the universe three
hundred thousand years after the big bang, reveals almost no structure
at all. (See Figure 1.)
The deviations in this temperature from exact smoothness (exact isotropy) are
measured in tens of {\it millionths} of a degree. However, those deviations
are important! They are the origin of all the complexity in the universe
we see today. As the universe evolves, these fluctuations grow,
collapse, and fragment through gravitational attraction
to become the galaxies, stars, and planets which
characterize the universe today. Initially very close to equilibrium, the
matter in the universe is thereby driven further from equilibrium. That
disequilibrium is necessary for chemistry, geology, life, biology,
and human history. 

The evidence of the observations then is that the universe was a simpler
place earlier than it is now --- more homogeneous, more isotropic, with
matter more nearly in thermal equilibrium. The aim of quantum cosmology
is a quantum
theory of this simple initial condition.

\section{Classical and Quantum Initial Conditions}
\label{sec:IV}

It is an inescapable inference from the physics of the last sixty years
that we live in a quantum mechanical universe --- a world in which the
basic laws of physics conform to that general framework for prediction we call
quantum mechanics. We perhaps have little evidence for peculiarly
quantum mechanical phenomena on large and even familiar scales, but there is
no evidence that the phenomena that we do see cannot be described in
quantum mechanical terms and explained by quantum mechanical laws. This
is the first reason that the search for a theory of the initial
condition is carried out in the framework of {\it quantum} cosmology.
There is, however, another reason: quantum indeterminacy is
probably necessary for a comprehensible basic,
scientific theory of the initial
condition.

To explain this necessity
 and also to understand a bit of the machinery of quantum
cosmology, consider a model universe. Suppose the universe
consists of a box the size of the visible universe
containing a large number $N$ of
particles interacting by fixed potentials. To simulate the expansion of
the universe we could let the box expand. That's actually not a bad
model for what goes on in more recent epochs of the universe.

Classically a history of this model universe is a curve in a $6N$
dimensional phase space of the positions and momenta of all the 
particles in the
box. Classical evolution is deterministic --- if the point 
in phase space specifying the system's configuration
 is known at one time, the location at all other times is
determined by the equations of motion. A classical
 theory of the initial condition
of the model universe
thus might specify the initial point in phase space at $t=0$. However, such a
theory would necessarily be hopelessly complex because it would have to
encode all the complexity we see today. Its description would be too
long to be comprehensible.

A {\it statistical} classical initial condition could be simpler. Such an
initial condition would only give a 
{\it probability} for the initial point in phase space and therefore only a
probability for the subsequent evolution. Present predictions
of the future would then be probabilistic. For example, 
observers at any time in the history of the universe 
can only see galaxies within a
distance close enough that their light could have reached them 
in the time
since the big bang. This cosmological horizon expands as the universe
ages. One new galaxy comes
over this cosmological horizon approximately 
every 10 minutes. A statistical
initial condition might not predict with near certainty, say, the specific
locations of the individual new 
galaxies, but rather their statistical distribution on the sky. Similarly, with
a classical initial condition in which matter was initially in thermal
equilibrium, one might predict the overall intensity of the background
radiation on the sky, but not the location of any particular fluctuation
in its intensity.

Probabilities in classical physics reflect ignorance. A classical
statistical law of the initial condition would mean that we have
some information about how the universe started out, but not all.
However, we learn from observation. With every observation we could
refine our theory of the initial condition which would therefore become
increasingly complex, reflecting the complexity of the present, and 
thus become increasingly less
comprehensible. 

Quantum mechanics is inherently indeterministic and probabilities are
basic. The most complete specification of the initial state of our model
box of particles would be a wave function on the configuration space of
all their positions
\begin{equation}
\Psi\left(\vec x_1, \cdots, \vec x_N\right)
\label{fourone}
\end{equation}
--- a wave function of the universe for this model. Unlike classical physics,
subsequent observation will not improve this initial condition, although
the results of observation can be used to improve future predictions. 
Thus, in quantum mechanics, it is
natural to have a simple, comprehensible law of the initial condition
which is consistent with the complexity observed today.\footnote{ For
an early statement of this, see \cite{Woosum}.}

\section{What is Predicted in Quantum Cosmology?}
\label{sec:V}

My colleague, Murray Gell-Mann, once asked me, ``If you know the wave 
function of
the universe, why aren't you rich?'' The answer is that
very little is predicted with certainty by such a quantum 
initial condition of the universe and certainly not of much use in
generating wealth. What might be  predicted by an initial
condition for cosmology is the subject of this Section.

Quantum mechanics predicts probabilities for sets of alternatives. 
In our model universe in a box, for example, it might predict the probabilities
for alternative ranges of
the position of a particle at a particular time, or the probabilities 
 for alternative distributions of energy density in the box, and many
other sets of alternatives. These are the probabilities for alternatives
which are {\it single}
events in a {\it single} closed system --- the universe as a whole. 

What do such probabilities of single events mean? Some may 
find it helpful to think of these probabilities as predictions of
relative frequencies in an imaginary infinite ensemble of universes, but
they are not frequencies in any accessible sense. Rather, to understand what
the probabilities of single events
mean it is best to understand how they are used. 
Probabilities of single events can be useful guides to behavior even
when they are distributed over a set of
alternatives so that none is very close to zero or one. 
Examples are the probability that it
will rain today or the probability of a successful marriage. However,
because the probabilities are distributed, the event which occurs ---
rain or no rain, divorce or death before parting  --- 
does not test the theory that produced the
probabilities. Tests of the theory occur when the probabilities
 are {\sl near certain}, by which I mean sufficiently close to zero or
one that the theory would be falsified if an event with probability
sufficiently close to zero occurred, or an event
with a probability sufficiently close to one did not
occur.\footnote{{\it How} close to zero or
one probabilities must be for near certain 
predictions
depends on the circumstances in which they are used
as I have discussed elsewhere \cite{Har91a}.} Various strategies can be used to
identify sets of alternatives  for which probabilities are near zero or one.
The most familiar is to study the frequencies  of outcomes of repeated
observations in an
ensemble of a large number of identical situations. Such frequencies would 
be predicted with certainty in an infinite ensemble. However, since
there are no genuinely infinite ensembles in the world, we are
necessarily concerned 
with the probability for the deviations of the frequency in 
a finite ensemble from the expected behavior of an infinite one. Those are 
probabilities  for  single properties  (the deviations) of a single system
(the whole ensemble) that become closer and closer to zero or one as
the ensemble is made larger. 

Another strategy to identify alternatives with 
 probabilities near zero and one is to
consider probabilities conditioned on other information besides that
given in the theory of dynamics and the
initial condition of the universe. Present theories of the initial
condition do not predict the observed orbit of Mars about the sun with
any significant probability. But they do predict that the {\it conditional}
probability for the observed orbit is near one {\it given} a few previous
observations of Mars' position.  Such conditional
probabilities are what are used in the rest of the sciences when they are 
viewed from the perspective of quantum cosmology as we shall discuss in
more detail in the subsequent Sections.

In the following discussion it will be helpful to use just a little of
the mathematics of quantum mechanics to discuss quantum cosmology.\footnote{
For more details at an elementary level see \cite{Har93a} and in greater
depth see \cite{GH93a}.} For
simplicity and definiteness let us continue to discuss the model
universe of $N$ particles
 in a box. The quantum initial state of this model universe is
represented by a state vector $|\Psi\rangle$ in a Hilbert space, or
equivalently 
by a wave function of the co\"ordinates of all the
particles in the box:
\begin{equation}
\Psi\left(\vec x_1, \cdots, \vec x_N\right)\ .
\label{fiveone}
\end{equation} 

General alternatives at a moment of time whose probabilities we might
want to consider  can always be reduced to a set of
``yes-no'' alternatives. For instance, questions about the position of
a particle can be reduced to questions of the form: ``Is the particle
in this region  --- yes or no?'', ``Is the particle in that region --- yes
or no?'', etc. 
A set of ``yes-no'' alternatives at one moment of time, say $t=0$, is
represented by a set of orthogonal projection operators $\{P_\alpha\}$,
$\alpha=1,2,\cdots$ --- one projection operator for each alternative.
(A projection operator is one whose square is equal to itself.)
The projection operators satisfy
\begin{equation}
\sum\nolimits_\alpha P_\alpha = I\ , \qquad {\rm and} \qquad P_\alpha
P_\beta =0\ ,\ \alpha \not= \beta\ ,
\label{fivetwo}
\end{equation}
showing mathematically that they represent an exhaustive set of
exclusive alternatives. The same set of alternatives at a later time $t$
is
represented by a set of (Heisenberg picture) projection operators
$\{P_\alpha(t)\}$. The time dependence of each $P_\alpha (t)$ is given by
\begin{equation}
P_\alpha(t) = e^{iHt} P_\alpha e^{-iHt}
\label{fivethree}
\end{equation}
where $H$ is the Hamiltonian encapsulating the basic dynamical theory.
The probability predicted for alternative $\alpha$ at time $t$ is
\begin{equation}
p(\alpha) = \left\Vert P_\alpha (t) | \Psi \rangle \right\Vert^2\ ,
\label{fivefour}
\end{equation}
where $\Vert\cdot\Vert$ means the length of the Hilbert
space vector inside. For this model, the Hamiltonian $H$ specifies the
first of the two elements of a basic physical theory described in
Section III --- the fundamental theory of dynamics. The state vector
$|\Psi\rangle$ or equivalently the wave function
$\Psi(\vec x_1, \cdots \vec x_n)$ specifies the second element --- 
the initial condition. 

Probabilities for alternatives at a moment of time are not the most
general predictions of quantum mechanics. More generally, one 
can ask for the probabilities of
sequences of sets of alternatives at a series of different times $t_1< 
t_2 <\cdots <
t_n$ making up a set of alternative {\it histories} for the universe.
Each history corresponds to a particular sequence of
alternatives $(\alpha_1,\cdots, \alpha_n)$ and
 is represented by an operator that is the chain of 
 projections corresponding to the sequence of alternatives
\begin{equation}
C_\alpha = P^n_{\alpha_n}(t_n) \cdots P^1_{\alpha_1}(t_1)\ .
\label{fivefive}
\end{equation}
Here, the index $\alpha$ is shorthand for the whole sequence $(\alpha_n,
\cdots, \alpha_1)$ and the superscripts on the $P$'s indicate that different sets of
alternatives can
be considered at different times. When the operator $C_\alpha$ is applied to 
the initial state vector $|\Psi\rangle$, one obtains the {\sl branch state
vector} $C_\alpha|\Psi\rangle$ 
corresponding to the history $\alpha$. The probability of the  
history $\alpha$ is the length of the history's branch state vector:
\begin{equation}
p(\alpha) = \Vert C_\alpha | \Psi \rangle\Vert^2\ .
\label{fivesix}
\end{equation}
Probabilities of histories are essential for predicting such everyday
things as the orbit of the moon, which is a sequence of positions at a
series of times.

We can now begin to analyze the question of what is predicted in quantum
cosmology and what is not predicted. The most characteristically quantum
mechanical limitation on what can be predicted is that not every set of alternative histories
that may be described can be assigned probabilities by the theory
because of quantum mechanical interference. That is very clearly
exemplified in the two-slit thought experiment illustrated in Figure 2.
Electrons proceed from
an electron gun through a barrier with two slits on their way to
detection at a screen. Passing through slit $A$ or slit $B$
defines two alternative histories for the electrons arriving 
at a fixed point $y$
on the screen. In the usual story if we have not measured which slit an
electron passed through, then it would be inconsistent to predict
probabilities for these alternative histories.
 It would be inconsistent because the
probability to arrive at $y$ would not be the sum of the probability to
pass through $A$ to $y$ and the probability to pass through $B$ to $y$:
\begin{equation}
p(y) \not= p_A(y) + p_B(y)\ .
\label{fiveseven}
\end{equation}
That is because in quantum mechanics probabilities are the squares of
amplitudes and
\begin{equation}
|\psi_A(y) + \psi_B(y)|^2 \not= |\psi_A(y)|^2 + |\psi_B(y)|^2\ .
\label{fiveeight}
\end{equation}
\begin{figure}[t]
\centerline{\epsfysize=3.00in \epsfbox{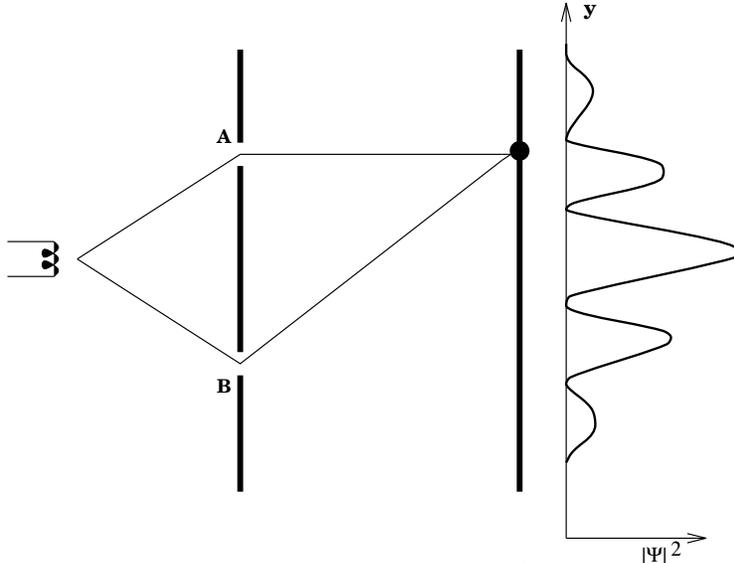}}
\caption
{\sl The two-slit
experiment.  An
electron gun at left emits an electron traveling towards detection at a
screen
at right,
its progress in space recapitulating its evolution in time. In between
there is a barrier with two slits. Two possible histories of an electron
arriving at a particular point on the screen are defined by whether it
went through slit $A$ or slit $B$.
In quantum mechanics,
probabilities cannot be consistently assigned to this set of two
alternative
histories because of quantum mechanical interference between them.
However, if the electron interacts with apparatus that
measures which of the slits it passed through, then interference is
destroyed, the alternative histories
decohere, and probabilities can be assigned to the alternative histories.}
\end{figure}

It is not that we are ignorant of which slit an electron passes through,
so that the probabilities are 50--50. It is inconsistent to discuss
probabilities at all.

Quantum mechanics, in any of its various levels of formulation,
therefore contains a rule specifying which sets of alternative histories
may be assigned probabilities and which may not. In the most general
 context of the
quantum mechanics of the universe that rule is as follows
\cite{Gri84,Omnsum,GH90a}: Probabilities
may be consistently assigned to just those sets of histories for which there
is vanishing interference between the individual members of the set as a
consequence of the universe's initial state $|\Psi\rangle$. Such sets of histories are
said to decohere. The condition for a decoherent set of
histories is that the branches of the initial state $C_\alpha|\Psi\rangle$ 
corresponding to
individual histories be mutually orthogonal:
\begin{equation}
\langle \Psi | C^\dagger_\alpha \cdot C_\beta | \Psi \rangle \approx 0
\ , \quad \alpha \not=\beta\ .
\label{fivenine}
\end{equation}
The most general probability sum rules are satisfied as a consequence.
Consistency limits the predictions of quantum theory to the
probabilities of {\it decoherent} sets of alternative histories. 

As an example of how the decoherence of a set of histories 
comes about, 
think about a single millimeter-sized dust grain in a quantum state that is a
superposition of two positions about a millimeter apart located 
deep in intergalactic space. Consider alternative histories of the
position of this particle at a sequence of a few times. 
(The $P$'s in (\ref{fivefive}) would then be 
projections onto ranges of this position.) Were the particle isolated,
this situation would be analogous to the two-slit experiment, and
histories of differing positions would not decohere.  However,
  even deep in space this
particle is not
isolated. The all-pervasive light from the big bang illuminates the
particle,
and about $10^{11}$ cosmic background photons scatter from it
every second. Through these interactions, this seemingly isolated dust
grain becomes correlated with radiation in a part of the universe whose size is
growing at the speed of light. The two states with different positions become
correlated with two different, nearly orthogonal
 states of the radiation after a time of
about a nanosecond. By this means, a branch of the initial state in which the grain
is initially at one position becomes  orthogonal to a branch in 
which the grain is a
millimeter away. Decoherence of alternative histories of position
has been achieved because the relative
phase between states of different position has been dissipated by feeble
interactions with the background radiation.
Mechanisms such this are widespread in the universe and
typical of those effecting the decoherence of
histories of the kinds of classical
variables we like to follow. (See, {\it e.g.}~\cite{JZ85,Zursum})

In the above example, decoherence of alternative histories 
of the position of the dust grain is achieved at the cost of
ignoring the photons that are effecting the decoherence. That is an
example of {\sl coarse-graining}. Were we to
consider a set of alternative histories of states of the cosmic
background 
radiation as well as the position of the grain we would be, in effect,
following all possible phase information. Such a set of alternative
histories would generally not decohere. Except for trivial cases, 
sets of histories must describe {\it coarse-grained}
alternatives in order for probabilities to be predicted at all. This
necessary imprecision is a genuine limit to what can be predicted in
quantum cosmology, in contrast to the limits of the kind associated with
the uncertainty principle which are merely limits to  the applicability
of classical modes of description.\footnote{Decoherence also implies
another kind of limit to classical predictability which should be
mentioned although we cannot discuss it in any depth here.  As
described, realistic mechanisms of decoherence involve the dispersal of
phase information concerning a subsystem into an environment that
interacts feebly with it. Those interactions produce noise which limits the
classical predictability of the subsystem. 
Thus, for classical predictability appropriate
and sufficient coarse-graining is needed for the decoherence necessary
to predict probabilities at all. But further coarse-graining is 
needed for the subsystem to have
sufficient inertia to resist the noise that those mechanisms of decoherence
produce and thereby  become classically predictable. (For an introductory
discussions see \cite{Har94b}. For a more detailed one see
\cite{GH93a}.)} 

We thus have the picture of a vast class of all possible sets of
alternative histories and a smaller subclass of decoherent sets of histories for
which quantum theory predicts probabilities. For almost none of these
decoherent
sets is there a history predicted with certainty on the basis of the
initial state alone. If one history has probability one, then all alternatives
to it must have probability zero. Suppose we have such a set, and let
$\alpha_c$ be the label of the certain history, then from
(\ref{fivesix}) 
\begin{equation}
\Vert C_\alpha |\Psi\rangle\Vert^2 = 0\ ,\quad \alpha\not=\alpha_c\ ,
\label{fiveten}
\end{equation}
which implies
\begin{mathletters}
\label{fiveeleven}
\begin{equation}
C_\alpha |\Psi \rangle =0\ , \quad \alpha\not=\alpha_c\ .
\label{fiveelevena}
\end{equation}
Then, since $\Sigma_\alpha C_\alpha = I$ as a consequence of (\ref{fivetwo}),
 we also have
\begin{equation}
C_\alpha | \Psi \rangle = |\Psi \rangle\ , \quad \alpha=\alpha_c\ .
\label{fiveelevenb}
\end{equation}
\end{mathletters}
Decoherence , eq.~(\ref{fivenine}), is then automatic for such sets of
histories in which one is certain.

Eq.~(\ref{fiveelevenb}) shows that 
operators of histories that are predicted with certainty act as
projection operators on the initial state. An alternative predicted with
probability one is thus mathematically equivalent to the alternative
corresponding to the question ``Is the universe in state
$|\Psi\rangle$''? 
These are very special questions. Out of the class of
sets of decoherent histories almost none correspond to sets in which one
history is a certain prediction of the initial condition and the theory
of dynamics alone. 

In quantum cosmology we might hope that  some
gross features of the universe might be among those that are 
predicted with near certainty
from the initial condition and dynamics alone. These include
features such as the approximate homogeneity and isotropy of the universe
on scales above several
hundred megaparsecs\footnote{A megaparsec (Mpc) is a convenient unit for
cosmology. One megaparsec = 3.3 million light years = $3.1\times10^{24}$cm.
The size of the universe visible today is of order several thousand
megaparsecs.}, its vast age after the big bang when measured on elementary
particle time scales, and certain features of the spectrum of density
fluctuations that grew to produce the galaxies. On more familiar scales we
may hope that the laws of the initial condition and dynamics would
predict the homogeneity of the thermodynamic arrow of time and the wide
range of scale and epoch on which the regularities of classical physics
are exhibited. There has even been speculation that phenomena on very small
scales, such as the dimensionality of spacetime or certain effective
interactions  of the elementary particles at accessible energy scales, 
may be near certain predictions of the initial condition
and dynamics. But there is little reason to
suspect that a simple theories  of the initial condition and 
fundamental
dynamics will predict anything about the behavior of the New York stock
market with near certainty and a great many other interesting phenomena
as well. That is why you can't get rich knowing the
wave function of the universe!

The situation is very different if information beyond laws
of dynamics and the initial condition is supplied and probabilities conditioned on
that information are considered.  There are many sets of {\it
conditional} probabilities in which one member of the set is near
certain. These conditional probabilities are the basis of prediction in
all the other sciences when viewed from the perspective of quantum
cosmology as will be described in the next Section.

I have described various limitations on what can be predicted in quantum
cosmology. Yet there is a sense in which we, as information gathering
and utilizing physical systems, make use of only a small part of the
possible predictions of quantum cosmology. That is because of our almost
exclusive focus
on alternatives defined in terms of the variables of classical physics 
--- averages over suitable volumes of densities of energy and momentum, densities of nuclear and chemical
species, average field strengths, etc. Such classical quantities are
represented by quantum operators called quasiclassical operators. (
They are termed {\it quasi}classical because they do not behave classically in all
circumstances.) Certainly our immediate experience can be described in
terms of quasiclassical variables even when --- as in the clicks of a Geiger
counter --- these variables do not obey deterministic classical laws.

Even in our theorizing about regions
of space or epochs in time that are very distant from us, we often focus on
histories of alternatives of quasiclassical operators. Only in the
microscopic arena do we consider non-quasiclassical alternatives such as
election spin and coherent superpositions of position. Even then
we typically consider such alternatives 
only when they are tightly correlated with a quasiclassical variable as
in a measurement situation.

However, quantum field theory exhibits many more kinds of variables 
than the small set of quasiclassical ones. Decohering sets of histories
can be constructed from alternative values of non-quasiclassical
operators as well as from quasiclassical ones. Indeed, the
quasiclassical sets of histories are but a small subset of the whole
class of decohering histories. Quantum theory does not  privilege one
set of decohering histories over another. Probabilities are predicted
for all such sets of alternatives.
Histories of non-quasiclassical alternatives are not beyond reach.  
Suppose we were to
make measurements of peculiarly quantum mechanical variables involving
large numbers of particles in regions of macroscopic dimensions. 
The histories that would
be relevant for the explanation of the outcomes of these measurements
would not be histories of quasiclassical variables in these regions, 
but rather
histories of the non-quasiclassical alternatives
that were measured. 
The reason for our preference for quasiclassical sets of alternative
histories, like all other questions concerning ourselves as particular
physical systems, probably lies in our evolutionary history ---
not in the framework of quantum theory itself.

\section{Differences Between the Sciences}
\label{sec:VI}

Using the conditional probabilities of quantum cosmology, a
 particular orbit of the earth about the sun could be predicted with near
certainty given a few previous positions of the earth and a description
of the earth and solar system in terms of the fundamental fields which
are the language of quantum cosmology. The probabilities for the outcome
of chemical reactions become near certain predictions of quantum
cosmology given a description in terms of fundamental fields of the 
molecules involved and the
conditions under which they interact. The
probabilities for the behavior of sea turtles in particular environments
could, in principle, become predictions of quantum cosmology 
given a description
of sea turtles and their environments in the language of quantum
cosmology. Even the probabilities for the different behaviors of human
--- beings  both individually and collectively --- could in principle be
predicted given a sufficiently accurate description of the individuals, 
their history, their environment, and their possible modes of behavior. In
this way {\it every} prediction in science could be viewed in terms of
a conditional probability in quantum cosmology. Why then do we have
separate sciences of astronomy, chemistry, biology, psychology, and so on?
The answer, of course, is that it is neither especially interesting 
nor practical to reduce the
predictions of these sciences to a computation in quantum cosmology.

One measure of the difference between the sciences is how sensitive the 
regularities they study are to the forms of the
 initial condition of the universe
and the fundamental theory of dynamics.
The phenomena studied in chemistry, fluid mechanics,
geology, biology, psychology, and human history, depend only 
very little on the particular form of
the initial condition. 
All of these sciences, especially chemistry, 
depend on the form of the theory of dynamics in some approximation, but as we
move through the list we are moving towards in the direction of the study of
the regularities of {\it increasingly specific subsystems of the universe}.
Specific subsystems can exhibit more regularities than are implied
generally by the laws of dynamics and the initial condition. 
The explanation of these regularities
lies in the origin and evolution of the specific subsystems in question. 
Naturally these regularities are more sensitive to 
this  specific history than they are to the form of the initial
condition and dynamics. That is especially clear in a science
like biology. Of course, 
living systems conform to the laws of physics and chemistry, but their detailed
form and behavior depend much more on the frozen accidents of 
several billion years of evolutionary history on a particular
planet moving around a particular star  than they do on the details of 
superstring theory or
the ``no-boundary'' initial condition of the universe.\footnote{See, e.g. 
\cite{Gel94} for more discussion in greater depth and examples from this point of view. }
Conversely the phenomena studied by these 
sciences do not help much in discriminating  among different theories
of the initial condition and dynamics. It is for such reasons that
it is not of pressing interest --- either for other areas of science or
for quantum cosmology itself --- to express the predictions of such 
phenomena 
as quantum cosmological probabilities, even though it is in principle possible
to do so.

Even if we were to wish to carry out a calculation of the conditional
probabilities in quantum cosmology necessary for prediction in the other
 sciences, an examination of what it would take 
yields three measures which distinguish the other sciences from quantum
cosmology and from one another. To yield a conditional probability the
theory requires:

\begin{itemize}
\item A description of the coarse-grained alternatives whose
probabilities are to be predicted in terms of fundamental quantum
fields.
\item A description of the circumstances on which the probabilities are
conditioned in terms of fundamental quantum fields.
\item A computation of the conditional probabilities.
\end{itemize}
\begin{table}[t]{}
\caption{Some Differences Between the Sciences}
\vskip .13 in
\begin{tabular}{l|l|l|l}
\qquad\qquad\qquad\qquad\qquad&Length of Coarse-grained&Length of
Coarse-grained
&Length of Computation of\\
&Description of Alternatives &Description of Conditions&Conditional
Probabilities\\ \hline\hline
Classical Physics&\qquad Very Short&\qquad Short&\qquad Very Short \\
\hline
Astronomy&\qquad Short&\qquad Short& \qquad Short --- Long\\ \hline
Fluid Mechanics &\qquad Short --- Long &\qquad Short &\qquad Short ---
Long\\ \hline
Chemistry&\qquad Short --- Long &\qquad Short --- Long &\qquad Long ---
Very Long\\ \hline
Geology&\qquad Long&\qquad Long& \qquad Long\\ \hline
Biology&\qquad Long --- Very Long&\qquad Long --- Very Long &\qquad Long ---
Very Long \\ \hline
Psychology &\qquad Very Long &\qquad Very Long &\qquad Very, Very
Long(?)\\ 
\end{tabular}
\end{table}
\vskip .13 in

The table above shows some simplistic guesses of the lengths of
these three parameters for typical problems in the various sciences. We can 
discuss a few of these:

By classical physics I simply mean Newton's laws of mechanics and
gravity, the laws of continuum mechanics, Maxwell's electrodynamics, the
laws of thermodynamics, etc. --- in short, the basic laws of physics as
they were formulated in the 19${\rm th}$ century. (I do not mean some
specific application of these laws, 
as to the breaking of ocean waves.) Classical physics might almost be
counted as a science separate from physics, for the laws of classical physics do not
hold universally, but only for certain kinds of subsystems in particular
circumstances.  However, the table shows the reason these laws are
usually considered part of the science of physics. There is just a short
list of quasiclassical variables (volume averages of fields, densities
of energy, momentum, chemical composition, etc.) whose ranges of values define the
coarse-grained alternatives of classical physics \cite{GH93a}.  It is a
somewhat longer business to spell out, in quantum mechanical terms, the
circumstances in which classical physics applies.  But the derivation of
the laws of classical physics can be as short as a journal 
paper.\footnote{For a one-journal-paper
derivation from the quantum cosmological point of view see, 
{\it e.g.}~Ref.~\cite{GH93a}.}

As we move down the table to astronomy we encounter more specific
classes of physical systems --- stars, clusters, galaxies, etc. However,
the difficulty of obtaining data on such distant objects prevents us from
learning much individual detail.  The length of the coarse-grained
descriptions of both conditions and alternatives are typically short.
The computations utilizing  the equations of classical physics, however, range
from very short dimensional estimates to long simulations of supernovae
explosions.

In fluid mechanics we encounter a wide variety of particular phenomena
arising from differential equations of classical physics. One has only
to mention laminar flow, turbulence, cavitation,
percolation, convection, solitons, shock waves, detonation,
superfluidity, clouds, dynamos, internal waves, ocean waves, the
weather, etc., to recall something of the richness of phenomena studied in
this subject. The coarse-grainings describing the alternative behaviors
of fluids can sometimes be long although the description of the
conditions is usually shorter. Many of these phenomena can be simulated
on computers today by solving the differential equations of classical
physics. These calculations could be considered calculations in
quantum cosmology were we 
to append to them a standard description of the
alternatives and conditions, together with the computations that 
justify the use of these approximate equations in terms of
the fundamental theory of quantum fields and the initial condition.

The description of the molecules of interest in chemistry can vary from
short --- as in typical chemical formulae ---
 to long  --- as in the base sequence in
human DNA. There is a similar range of conditions for chemical reactions
ranging from a few reagents in a test tube to the interiors of cells.
Quantum chemists {\it can} compute certain chemical properties such as
the those of chemical bonds directly from the equations of an effective
low-energy theory of the elementary particles but these computations can
only be described as long.

In geology we have a science concerned with a very specific system ---
the earth --- observed in considerable detail. A lengthy string is
needed to describe the alternative configurations and composition of the
material on the surface in the detail that we know it. A long history would
have to be described to set the conditions for calculating the
probabilities and calculations of these probabilities, even assuming the
laws of classical physics, would be very long.

The reader probably needs little convincing that the description of the
behavior of a complex biological organism plus its evolutionary history
and its
present environment in the language of quantum field theory would be a
long business indeed! We should not pretend that we are anywhere
close to being able to give such a description or 
to being able to carry out the relevant computations of
conditional probabilities in quantum cosmology. Psychology and human history
are yet more
difficult. We may have a rough idea of how to describe the action of a
bird's beak in the language of quantum field theory, but very little idea
of the coarse grainings that describe an individuals thoughts  and emotions or the vissitudes
of empires. 

Dear reader, please do not write the author concerning the inadequacies
of the above discussion.  He is aware that the boundaries between the
sciences are not precisely defined and that there is wide variation in
these three parameters within each one. In astronomy, for example, 
the description of
our nearest star --- the sun --- can be just as complex as that of any
phenomena in fluid mechanics (and indeed is a part of fluid mechanics). 
The smallest self-reproducing biological
units may be simulable by conceivable computers \cite{Fra95}. There may be
universal principles of mind which derive rather directly from the
basics of physics \cite{She94}. The important point is, that at a
basic level, every prediction in science may be viewed as the
prediction of a conditional probability for alternatives in quantum
cosmology and that
the probabilities relevant to
different sciences may be distinguished, in part, by 
their sensitivity to the theories of the initial condition and dynamics,
by the 
length of the description of the alternatives, by the length of the
description of the conditions, and by the length of the computation needed
to
produce them. 

\section{Limits to Implementation}
\label{sec:VII}

The preceding Section discussed some limitations of practice in our effort to
implement the predictions of quantum cosmology for interesting specific
subsystems in the universe. These limits were of the general character
described in Section IIB. Are there more fundamental and general limits
arising from computational intractability? Quantum cosmology provides
some examples.

There are physical reasons for computational intractability and
mathematical ones. Landauer \cite{Lansum} has raised the issue of
whether there are predictions whose computation would require more resources
of space, material, and time than are available in the universe.
Quantum cosmology may also present an example of what might be regarded
as an extreme example of mathematical computational intractability. 
There is some evidence that the wave function of the universe might be
non-computable in the technical mathematical sense. 

One idea for a
theory of the wave function of the universe is the ``no-boundary''
proposal \cite{HH83}.  To understand a little of this idea assume, 
for simplicity,
that the universe is spatially closed and that 
gravity is the only quantum field. 
A cosmological wave function is then a
function of the the possible geometries of three-dimensional space. The
``no-boundary'' idea is that the value of the wave function of the universe
$\Psi$
at one particular spatial geometry is a sum over all locally
Euclidean four-dimensional geometries which have this three-dimensional
space as a boundary and {\it no other boundaries}. Each four-dimensional
geometry ${\cal G}$ in the sum
is weighted by $\exp\,(-I[{\cal G}])$ where $I[{\cal G}]$ is the
classical action for the geometry. Mathematically a geometry is a
specification of a notion of distance (a metric) on a space such that
any small region can be smoothly mapped into a 
to a region of flat Euclidean
space ( a manifold). A sum over geometries would therefore
naturally include a sum over manifolds as well as a sum over metrics. By
suppressing two of the four dimensions we can give a crude pictorial
representation of this double sum as shown in Figure 3. 

\begin{figure}
\centerline{\epsfysize=3.50in \epsfbox{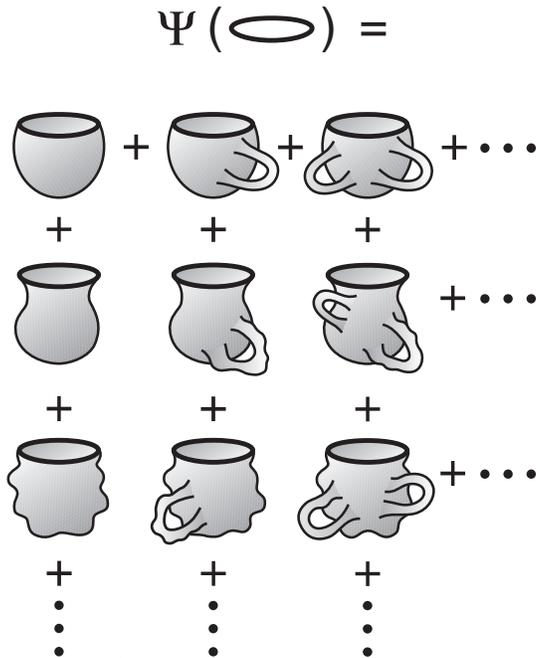}}
\caption      
{\sl The wave function of the universe as a sum over manifolds and metrics.
This figure uses two-dimensional analogs to illustrate some of the
ideas
that enter into the construction of the ``no-boundary''
wave function of the universe. That wave function is a function of
three-dimensional spatial geometries one of which is represented here in two
fewer
dimensions by the heavy circular curve. For that given three-geometry, the
``no-boundary'' wave function is a sum over Euclidean four-geometries
that have it as one boundary
and {\it no other boundary}.
This sum can be divided into a sum over four-manifolds and a sum over
different four-metrics on those manifolds. The two-dimensional analog of
this sum is shown above. The surfaces in each column represent different
metrics on the {\it same} manifold. The manifolds in each column are the
same because the surfaces can be smoothly deformed into one
another by changing their shape. The metrics are different from one
surface
to another in a given column because the distance between two
points is generally different from one shape to another.
For example, the overall surface area may differ from one shape to
another. The two-dimensional surfaces in different columns are different
manifolds because they have different numbers of handles, and
surfaces with different number of handles
cannot be smoothly deformed into one another. A sum over manifolds is
thus analogous to the sum over columns. A sum over metrics is analogous
to the sum over different surfaces in each column.}
\end{figure}

The mathematics of quantum gravity has not been developed to the
point that we have a precise mathematical formulation of what the
relation schematically represented in Figure 3 might mean. 
One idea for making it precise is to approximate
each term in the sum by a manifold constructed of flat four-simplices
--- the four-dimensional analogs  of triangles in two-dimensions and
tetrahedra in three-dimensions. 
The two-dimensional analog of such a simplicial manifold
would be a surface made up of triangles like
a geodesic dome as illustrated in Figure 4.
To calculate in four-dimensions the sum crudely pictured in 
Figure 3 one would proceed as follows:
Choose a large number of four-simplices $N$. Find all possible
manifolds that can be made by joining these four-simplices together. 
Choose
{\it one} such assembly to represent each manifold in the sum. 
Integrate $\exp\,(-I[{\cal
G}])$ over the edge lengths of the simplices that are 
compatible with the triangle and similar
inequalities to approximate the sum over metrics. Sum the result over all 
manifolds. Take the limit as
$N\to\infty$. That is one possible way the sum over geometries in the
``no boundary'' proposal for the wave function of the universe might be
implemented.\footnote{For more details and references to the earlier
literature see, {\it e.g.}~Ref.~\cite{Har85a}.}

A computer program to carry out this task would first have to try all possible
ways of assembling $N$ four-simplices together and reject those which do
not give a manifold. This is already a formidable mathematical problem
and it has only been recently proven  that an algorithm
exists to carry out this computation for four-dimensional manifolds
\cite{Rub92,Tho94}. The next step would be for the
computer to take this list of four-manifolds and eliminate duplications.
However, it is known that the issue of whether two simplicial 
four-manifolds are identical
 is undecidable.\footnote{For a review see \cite{Hak73}.} More
precisely, there does not exist a computer program which, {\it for any}
$N$, can compare two input assemblies of $N$ four-simplices making
up manifolds, and halt after having printed out ``yes'' if the manifolds are
identical and ``no'' if they are not.

This suggests that the wave function of the universe defined by a
sum over geometries that includes a sum over manifolds
is a non-computable number.\footnote{We are specifically assuming the
Turing model of computability.}  However, appearances can
be deceptive. Whether a number is non-computable or not is a property of
the number and not of the way it is represented. Merely exhibiting one
non-computable representation like the series in Figure 3 does not establish
that there is not some other representation in which it {\it is} computable.
Demonstrating non-computability in such cases is likely to be a
difficult mathematical problem.

\begin{figure}	
\centerline{\epsfysize=5.00in \epsfbox{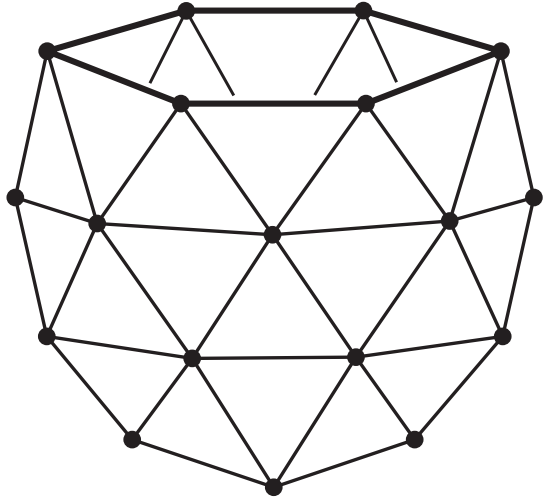}}
\caption
{\sl A smooth two-dimensional surface may be approximated by an assembly of
flat triangles like a geodesic dome. In an analogous way, a curved
four-dimensional geometry may be approximated by an assembly of
four-simplices --- the four dimensional analog of  two-dimensional
triangles or three-dimensional tetrahedra. The four-simplices must be
assembled so as to make a manifold --- a space such that any small
region can be smoothly mapped to a region of Euclidean space. Suppose
we are given $N$ four-simplices. Assembling them in different ways can
give different manifolds (the different
vertical columns in Figure 3). Different
assignments of lengths to the edges give different sizes and shapes on a
given manifold, that is, a different metric (the different shapes within
each column in Figure 3). A sum over geometries,
which is a sum over manifolds and metrics, may therefore be approximated
by choosing one assembly to represent
each  manifold in the sum, integrating over its
possible edge-lengths, and summing over all manifolds. The resulting
sum may be a non-computable number because there is no algorithm for
deciding when two simplicial four-manifolds are identical.}
\end{figure}

This suggested non-computability of sums over topologies has 
been taken as motivation for modifying the theory of the initial 
condition so that it is clearly computable \cite{Har85b,SW93}. 
But suppose that the wave function of the universe were non-computable. What
would be the implications for science? Bob Geroch and I 
 analyzed the
implications of non-computability for physics in 1986 \cite{GH86}. 
Our conclusion was that the
prediction of non-computable numbers would not be a disaster for
physics. That is because at any one time one needs theoretical
predictions only to an accuracy consistent with experimental possibilities.
Suppose, for example, it was sufficient for comparison with present
observations to know the wave function of the universe to an accuracy of
10\%. Suppose further it could be shown that to achieve this accuracy
only simplicial manifolds with less than 100 four-simplices need be
included in the series defining the wave function. 
The theorem concerning the non-existence of an algorithm for deciding the
identity of simplicial four-manifolds refers to an algorithm
that would work for {\it any} $N$. It
does not rule out establishing the identity of two four-manifolds with
less than 100 four-simplices. Indeed, being a problem that involves a
finite number of specific cases, one imagines it could be solved
with sufficient work on those cases. What the theorem ensures is that, if 
observations improve, and the
wave function is later needed to an accuracy of 1\%, requiring 
manifolds with a larger number of four-simplices  (say 10,000), a new
intellectual effort will be required to compute it.  
The algorithms that worked for
manifolds assembled from less than 100 four-simplices are unlikely to work for
manifolds assembled from less than 10,000 four-simplices.  

Thus, the prediction of non-computable numbers would not mean the end of
comparison between theory and observation. It would mean that the
process of computing the predictions could be as conceptually 
challenging a problem as posing the theory itself.

\section{Limits to Verification}
\label{sec:VIII}

Quantum mechanics predicts the probabilities of alternative histories of
the universe.  We cannot interpret these probabilities as predictions of
frequencies which are accessible to test, for we have access to but a single
universe and but a single history of it. Our ability to test the
theory or to infer the theory from empirical data is therefore limited.

An example of current interest is ``cosmic variance'' in the predictions
of temperature
fluctuations in the cosmic background radiation. The observed pattern
of temperature determines the  correlation function $C(\theta)$ between the
temperature fluctuations $\delta T$
at two different directions $\vec n_1$ and
$\vec n_2$ on the sky separated by an angle $\theta$:
\begin{mathletters}
\label{eightone}
\begin{equation}
C(\theta) = \left\langle\frac{\delta T(\vec n_1)}{T}
\ \ \frac{\delta T(\vec
n_2)}{T}\right\rangle\ , 
\label{eightone a}
\end{equation}
where $\langle \cdot \rangle$ denotes an average over all directions
$\vec n_1$ and $\vec n_2$ such that $\vec n_1 \cdot \vec n_2 = \cos
\theta$.
This correlation function can be conveniently expanded in spherical
 harmonics $P_\ell (\cos \theta)$:
\begin{equation}
C(\theta)= {\buildrel \infty \over {\mathop\sum\limits_{\ell=0}}}
\ \frac{2\ell+1}{4\pi}\ C_\ell P_\ell
(\cos\theta)\ .
\label{eightone b}
\end{equation}
\end{mathletters}
The coefficients $C_\ell$ so defined
are the way the data from observations are
are usually quoted and are the objects of theoretical prediction.\footnote{ 
See {\it e.g.}, 
Ref.~\cite{Bonpp} for a detailed review.}

The probabilities of temperature fluctuations in the cosmic background
radiation are
predicted from a spectrum of fluctuations implied by the
initial quantum state. The probabilities of these fluctuations
are thus a detailed prediction of quantum cosmology that stem directly
from the initial condition. They are not
conditional probabilities requiring other information. 
The theory does not predict high probabilities for particular
fluctuations in temperature at particular locations on the sky. Rather, it
predicts distributed probabilities for these fluctuations (See {\it e.g.}, 
Ref.~\cite{HHaw85}), or equivalently the probabilities for various values
of the $C_\ell$. 
The expected value and the
standard deviation of this distribution is shown in Figure 5.
The width of the distribution is ``cosmic variance''. 
\begin{figure}
\centerline{{\epsfysize=6.00in \epsfbox{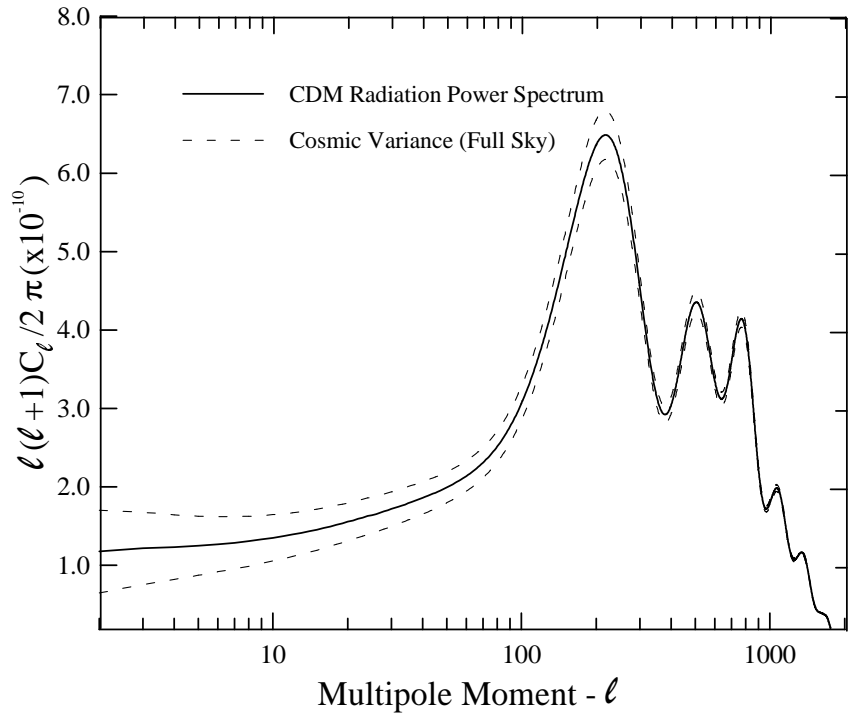}}}
\caption      
{\sl Cosmic Variance. The heavy line on this figure shows the expected value
of the multipole moments of the two point correlation function
defined by eq.~(8.1)
for temperature fluctuations in the cosmic background radiation
as predicted from the probabilities
of these fluctuations arising from a simple theory of the universe's
initial
condition. The
dotted lines show the standard deviation of the predicted distribution
called the  ``cosmic variance''. Observations of our single universe
yield
the correlation function and one particular distribution of observed
multipole moments. These observations will not distinguish two theories
of the initial condition whose ``cosmic variance'' both surround the
observed distribution. [Graph by J.~Gundersen.] }
\end{figure}

We cannot test these probabilistic predictions for the cosmic background
temperature fluctuations by measuring these fluctuations in a large
number of identical cases. We have only one universe and only one set of
observed temperature fluctuations!
An observed distribution of $C_\ell$'s 
inside this ``cosmic variance'' 
would be confirmation of the theory of the initial condition. An
observed distribution outside it would be evidence against it.  However,
observations will not distinguish two theories of the initial condition whose
``cosmic variance'' both surround the observed distribution. 
Thus, for such probabilistic  predictions we are inevitably limited in our ability to
test a theory of the initial condition.

More generally, as mentioned above, a theory of the initial condition can
be tested only through predictions whose probabilities are so near
certain that we would reject the theory if they were not observed. The
sets of histories which lead to near certain predictions are just a
small set of those for which probabilities are predicted.

There are limits, therefore, to the process of inferring the initial
state of the universe from observation. If $C_{\rm obs}$ is the operator
describing the entirety of our collective observations then strictly
speaking all we can conclude about an initial state $|\Psi\rangle$
is that it is not such that
\begin{equation}
C_{\rm obs} |\Psi\rangle=0\ .
\label{seventhree}
\end{equation}

That is not much of a restriction. For example, suppose that the
projection $P_{\rm pres.\ data}$ represents all our present data
including our records of the past history. It is not possible on the
basis of either present or future observations to distinguish an initial
state $|\Psi\rangle$ from that defined by
\begin{equation}
|\Psi^\prime\rangle= \frac{P_{\rm pres.\ data} |\Psi\rangle}{\Vert
P_{\rm pres.\ data} |\Psi\rangle\Vert}\ .
\label{sevenfour}
\end{equation}
Retrodictions of the past from present data and $|\Psi\rangle$
could differ greatly from those from the same data and 
$|\Psi^\prime\rangle$.\footnote{Unlike classical physics,
where the past can be retrodicted from sufficiently precise present data alone,
retrodiction in quantum theory requires present data {\it and} the initial
condition of the system in question. For further discussion see, {\it e.g.}
\cite{Har91a}, Section II.3.1.}  
But retrodictions are not accessible to
experimental check and therefore do not distinguish the two candidate
initial states. The two
initial conditions $|\Psi\rangle$ and $|\Psi^\prime\rangle$ could differ
greatly in complexity if the description of $|\Psi\rangle$ is short but
that of $P_{\rm pres.\ data}$ is long, and we may choose between these
physically equivalent possibilities on the basis of simplicity.
The search for a theory of the
initial condition must therefore rely on the principles of simplicity
and connection with the fundamental dynamical theory in an essential
way. 

Why is it that the fundamental dynamical theory --- the Hamiltonian of the
elementary particle system --- seems so much more accessible to
experimental test and so much easier to infer from observational data
than the theory of the initial condition? Strictly
speaking it is not. Were the Hamiltonian of the elementary particle
system to vary on cosmological scales --- to be a function of spacetime
position of the form $H(x)$ --- then inferring $H$ would be just as
difficult a process as inferring the initial $|\Psi\rangle$. However, we
{\it assume} the principle that the elementary particle interactions are local
in space and time. With that assumption the Hamiltonian describing
these interactions becomes accessible to many local tests on all sorts
of scales ranging from those accessible in particle accelerators to the
expansion of the universe itself. The problem of inferring the initial
$|\Psi\rangle$ therefore is not so very different from that of inferring
$H$ in making use of the theoretical assumptions. It is just that the
assumption of locality is so well adapted to the
quasiclassical realm of familiar experience that many more tests can be
devised on small scales of a theory of $H$
 than we are ever likely to find of a theory of
$|\Psi\rangle$ on cosmological scales.

\section{Conclusions: The Necessity of Limits to Scientific Knowledge}
\label{sec:IX}

If the world is complex and the laws of nature are simple, then there
are inevitable
limits to science. Not everything that is observed can be predicted;
only certain regularities of those observations can be predicted.  
Even given a theory, computational intractability or observational
difficulty may limit our ability to predict.
In a world of finite
observations, there are inevitable limits to our ability to discriminate
between different theories by the process of induction and test.

Quantum cosmology --- the most general context for prediction in science
--- exhibits examples of all three kinds of limits to scientific knowledge.
There are only a
very few predictions of useful probabilities that are conditioned
solely on simple theories of dynamics and the universe's  
initial condition. There is a far richer variety  of useful
probabilities conditioned on further empirical data that are the basis
for most of the predictions in science. There are some indications that
the ``no boundary''  initial wave function is non-computable in the
technical sense of yielding non-computable numbers. That does not limit
our ability to extract predictions from the theory in principle, but may
be an indication that predictions sensitive to the topological structure
of spacetime on small scales could be conceptually challenging
to compute.  Finally, it
is possible to exhibit different theories of the initial condition with
identical present and future predictions which can only be discriminated
between by an appeal to principles of simplicity and harmony with
fundamental dynamical laws.

We should not conclude a discussion of limits in science without
mentioning that science is useful {\it because} of its limits. Complex
adaptive systems are successful in evolution and individual
behavior because they identify and 
exploit the regularities that the universe
exhibits. Scientific theories predict what these regularities are and
explain their origin. Theories
can be used to estimate how tractable these predictions
 are to compute or practical
to measure. By comparing different theories induced from the same data
an idea can be gained of the reliability of our predictions. The existence
of limits of the kind we have discussed therefore
does not represent a failure of
the scientific enterprise. Limits are inherent in the nature of that
enterprise, and their demarcation is an important scientific question.

\acknowledgments

The author is grateful to M.~Gell-Mann for discussions of a number of
the issues discussed in Section VIII over many years. Thanks are also due to
H.~Morowitz and R.~Shepard for instruction in certain aspects of
biology and psychology respectively, and to J.~Gundersen for supplying 
several of the figures. This work was supported in
part by NSF grants PHY90-08502, PHY95-075065,  and  PHY94-07194.


\begin{references}

\bibitem{Boh34} N.~Bohr, {\sl Atomic Physics and the Description of
Nature}, Cambridge University Press, Cambridge, UK (1934).

\bibitem{Gel94} M.~Gell-Mann, {\sl The Quark and the Jaguar}, W.~Freeman
San Francisco (1994).

\bibitem{Hal91} J.~Halliwell,  in {\sl Quantum
Cosmology and Baby Universes:  Proceedings of the 1989 Jerusalem Winter
School for Theoretical Physics}, eds.~S.~Coleman, J.B.~Hartle, T.~Piran,
and S.~Weinberg, World Scientific, Singapore (1991) pp. 65-157.

\bibitem{Woosum} C.H.~Woo, {\sl Phys.~Rev.}, {\bf D39}, 3174, (1989),
{\sl Found. Phys.}, {\bf 19}, 57, (1989), and in {\sl Complexity,
Entropy, and the Physics of Information}, ed. by W. Zurek (Addison
Wesley, Reading, MA, 1990).

\bibitem{Har91a} J.B.~Hartle, in {\sl
Quantum Cosmology and Baby Universes:  Proceedings of the 1989 Jerusalem
Winter
School for Theoretical Physics}, ed. by ~S.~Coleman, J.B.~Hartle,
T.~Piran,
and S.~Weinberg, World Scientific, Singapore (1991) pp. 65-157.

\bibitem{Har93a}J.B.~Hartle, 
in {\sl Directions in General Relativity,
Volume 1: A Symposium and Collection of Essays in honor of Professor
Charles W. Misner's 60th Birthday}, ed. by B.-L.~Hu,
M.P.~Ryan, and C.V.~Vishveshwara, Cambridge University Press, Cambridge
(1993). gr-qc/9210006.

\bibitem{GH93a} M.~Gell-Mann and J.B.~Hartle, {\sl Phys.~Rev.},
{\bf D47}, 3345, (1993). gr-qc/9210010.

\bibitem{Gri84} R.~Griffiths, {\sl J.~Stat.~Phys.}, {\bf 36}, 219
(1984).

\bibitem{Omnsum} R.~Omn\`es, {\sl J.~Stat.~Phys.}, {\bf 53}, 893 (1988),
{\sl ibid}, {\bf 53}, 933 (1988); {\sl ibid}, {\bf 53}, 957 (1988); 
{\sl ibid}, {\bf 57}, 357 (1989); {\sl Rev.~Mod.~Phys.}, {\bf 64}, 339 
(1992); {\sl
Interpretation of Quantum Mechanics}, Princeton University Press,
Princeton, (1994).

\bibitem{GH90a} M.~Gell-Mann and J.B.~Hartle, 
in {\sl Complexity, Entropy,
and the Physics of Information, SFI Studies in the Sciences of
Complexity}, Vol.  VIII, ed. by W. Zurek,  Addison Wesley, Reading, MA
or in {\sl Proceedings of the 3rd
International Symposium on the Foundations of Quantum Mechanics in the
Light of
New Technology} ed.~by S.~Kobayashi, H.~Ezawa, Y.~Murayama,  and
S.~Nomura,
Physical Society of Japan, Tokyo (1990).

\bibitem{JZ85} E.~Joos  and H.D.~Zeh, {\sl Zeit.~Phys.}, {\bf B59}, 223
(1985).

\bibitem{Zursum} W.~Zurek, {\sl Phys.~Rev.~D}, {\bf 24}, 1516 (1981),
{\sl ibid.} {\bf 26}, 1862 (1982).

\bibitem{Har94b} J.B.~Hartle, 
in {\sl Proceedings of the Cornelius  Lanczos International
Centenary Conference}, North Carolina State University,
December 1992, ed.~by J.D.~Brown, M.T.~Chu, D.C.~Ellison, R.J.~Plemmons,
SIAM, Philadelphia, (1994). gr-qc/9404017; also J.B.~Hartle in the
Proceedings of the SFI conference on {\sl Fundamental Sources of
Unpredictability} to be held March 28--30, 1996.

\bibitem{Fra95} C.M.~Fraser et.~al., {\sl Science}, {\bf 270}, 397 (1995).

\bibitem{She94} R.N.~Shepard, {\sl Psychonomic Bulletin \& Review}, {\bf
1}, 2 (1994), {\it World and Mind} (to be published).

\bibitem{Lansum} R.~Landauer, {\sl IEEE Spectrum},  {\bf 4}, 105 (1967);
in {\sl Proceedings of the 3rd
International Symposium on Foundations of Quantum Mechanics in the Light
of New
Technology}, ed.~by S.~Kobayashi, H.~Ezawa, Y.~Murayama, and S.~Nomura,
Physical Society of Japan, Tokyo (1990); and 
{\sl Physics Today}, {\bf 44}, 23 (1990).

\bibitem{HH83} J.B.~Hartle and S.W.~Hawking, {\sl Phys.~Rev.}, {\bf
D28},
2960, (1983).

\bibitem{Har85a} J.B.~Hartle,  {\sl J.~Math.~Phys.}, {\bf 26}, 804,
(1985a).

\bibitem{Rub92} H.~Rubenstein, {\it The Solution to the Recognition
Problem for $S^3$}, unpublished lectures, Haifa, Israel (1992).

\bibitem{Tho94} A.~Thompson, {\sl Math.~Res.~Lett.}, {\bf 1} 613 (1994).

\bibitem{Hak73} W.~Haken, in {\sl Word Problems}, ed.~by W.W.~Boone,
F.B.~Cannonito and R.C.~Lyndon, North Holland, Amsterdam, (1973).

\bibitem{Har85b} J.B.~Hartle, 
{\sl Class. \& Quant. Grav.}, {\bf 2}, 707,
198{5b}.

\bibitem{SW93} K.~Schleich and D.~Witt,  {\sl
Nucl. Phys.}, {\bf 402}, 411, 1993;
{\sl ibid.}, {\bf 402}, 469, 1993.

\bibitem{GH86} R.~Geroch  and J.B.~Hartle, {\sl Found.~Phys.}, {\bf 16},
533 (1986).

\bibitem{Bonpp} J.R.~Bond, in {\sl Cosmology and Large Scale Structure,
Proceedings of the Les Houches School, Session LX, August 1993}, ed.~by
R.~Schaefer, Elsevier Science Publishers, Amsterdam (1995).

\bibitem{HHaw85} J.~Halliwell and S.W.~Hawking, {\sl Phys.~Rev.~D},
{\bf 31}, 1777 (1985).


\end{references}
\end{document}